\documentclass[final,dvipdfm,a4paper]{aipproc}
\layoutstyle{6x9}

\usepackage[dvipdfm]{hyperref} 
\usepackage{amssymb}           
\newcommand{\vv}[1]{\mathbf{#1}}       
\newcommand\sun{\hbox{$\odot$}}
\newcommand\degr{\hbox{$^\circ$}}
\newcommand{\Rvir}{R_\mathrm{vir}}
\newcommand{\Msol}{\mathrm{M_{\sun}}}   
\newcommand{\munit}{ \, h^{-1} \Msol }  
\newcommand{\Mpc}{\mathrm{Mpc}}         
\newcommand{\lunit}{ \, h^{-1} \Mpc }   
\newcommand{\tunit}{\, \mathrm{Gyr} }   
\newcommand{\jsc}{\tilde{j}}            

\begin{document}

\title[Spin Flips]{Spin Flips: Variation in the Orientation of Dark Matter
  Halos over their Merger Histories\footnote{The original poster is available at
    \url{http://www.astro.uni-bonn.de/\~pbett/pub.html}.  }}





\author{Philip E. Bett}{
  address={Argelander-Institut f\"ur Astronomie, Universit\"at Bonn, Auf
    dem H\"ugel 71, D-53121 Bonn, Germany},
}

\begin{abstract}
  Semi-analytic models of galaxy formation typically form the
  spheroidal components of galaxies (``bulges''), solely through
  galactic major mergers. However, it is possible that non-merger
  events (e.g. a ``fly-by'' by a smaller halo) can perturb a
  galaxy--halo system sufficiently to form a bulge.  We present a
  preliminary investigation into the frequency of major changes in
  halo and galaxy spin direction, which could be signatures of such
  events.
\end{abstract}

\maketitle



\section{Flip and Merger Event Distribution}\label{s:hMS}
Using a large cosmological dark matter simulation,\footnote{This is the ``hMS''
  simulation of a $100\lunit$ box; see e.g. \citep{Bett09} for details.}
we look at the evolution of the haloes identified at $z = 0$ by tracking their
properties through their most-massive progenitors at each simulation
output time $t_i$.  We call the change in the properties of a halo between
timesteps (e.g. from $t_{i-1}\rightarrow t_i$) an \emph{event}, and we
consider the distribution of events for all the selected\footnote{Each halo
  must have at least $1000$ particles (i.e. $M \gtrsim 9.5\times10^{10}
  \munit$), and have sufficiently large $\vv{j}$ that its orientation is not
  subject to discreteness bias: $\log_{10}\jsc \equiv
  \log_{10}\left(|\vv{j}|/\sqrt{G M \Rvir}\right) \geq -1.5$ (see
  \cite{Bett09}).}
   haloes in the $46$ timesteps from $z<6.2$.  In
particular, we compute the fractional halo mass change $\Delta\mu(t_i) = 1 -
M(t_{i-1})/M(t_i)$, and the angular momentum direction change $\cos\theta(t_i)
= \left(\vv{j}(t_i)\cdot\vv{j}(t_{i-1})\right) /
\left(|\vv{j}(t_i)||\vv{j}(t_{i-1})|\right)$, in terms of the halo mass
$M(t_i)$ and specific angular momentum vector $\vv{j}(t_i)$.  We show the
distribution of events in terms of $\Delta\mu$ and $\cos\theta$ in the left
panel of Fig.~\ref{f:spinflips}, and their cumulative distributions in the
middle and right panels.  These show that major mergers are much more likely
to than minor mergers to coincide with spin direction changes of more than a
given angle, $\theta_0$.  However, most events with significant direction change are
\emph{minor} mergers, not major.

\newcommand{\smlfig}{0.28\columnwidth}  
\begin{figure}
  \includegraphics[width=\smlfig]{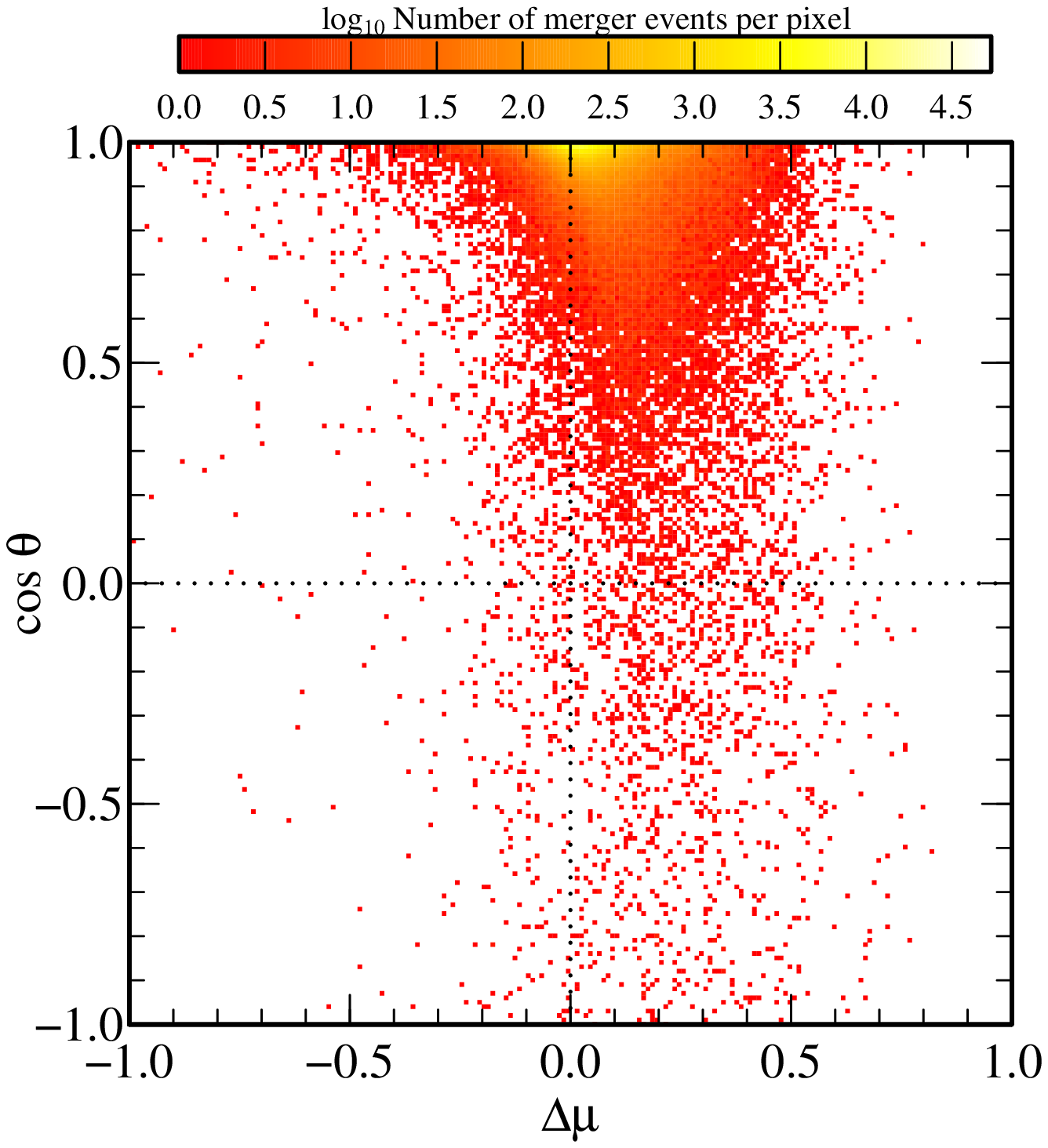}   
  \includegraphics[width=\smlfig]{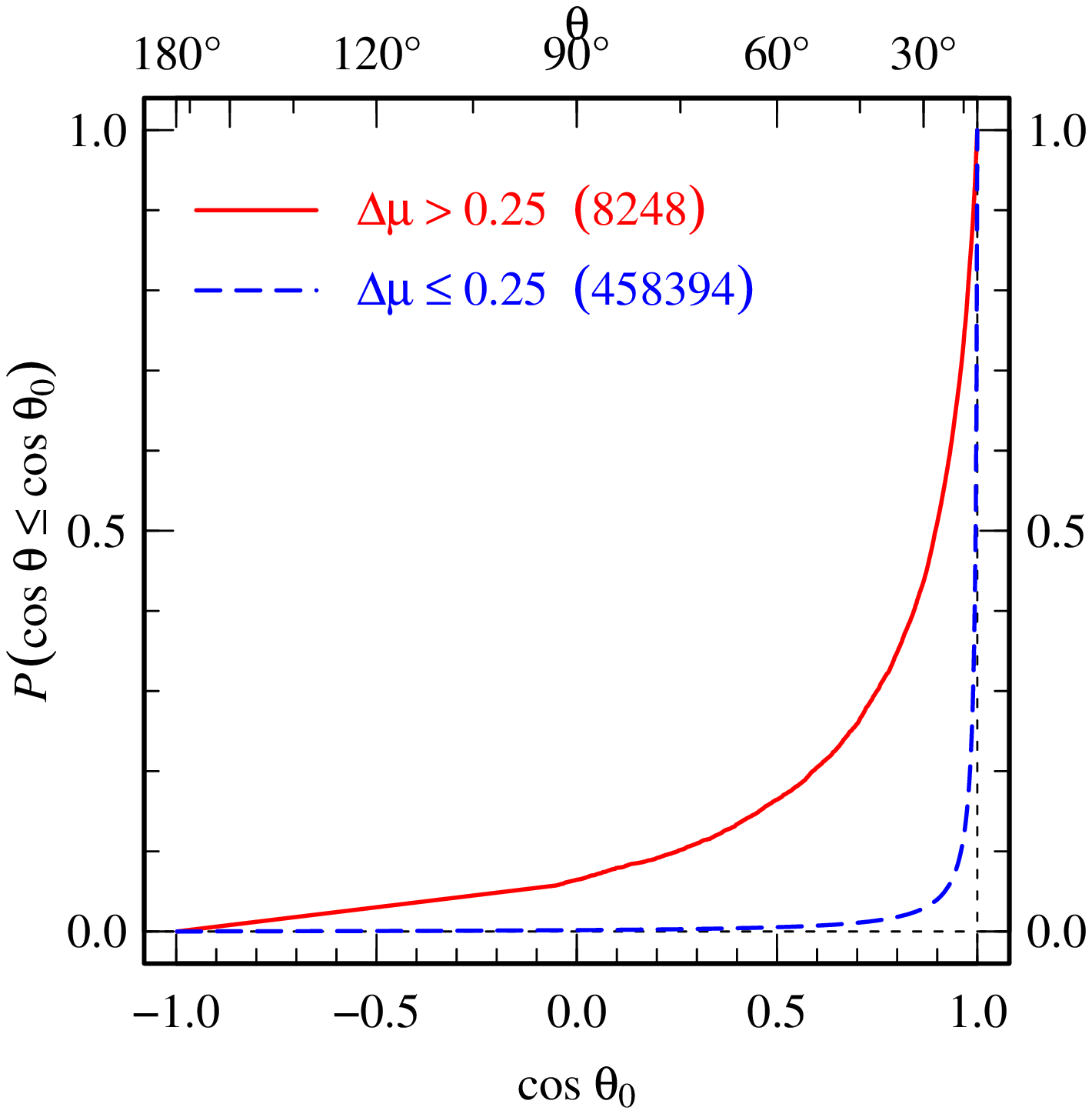}
  \includegraphics[width=\smlfig]{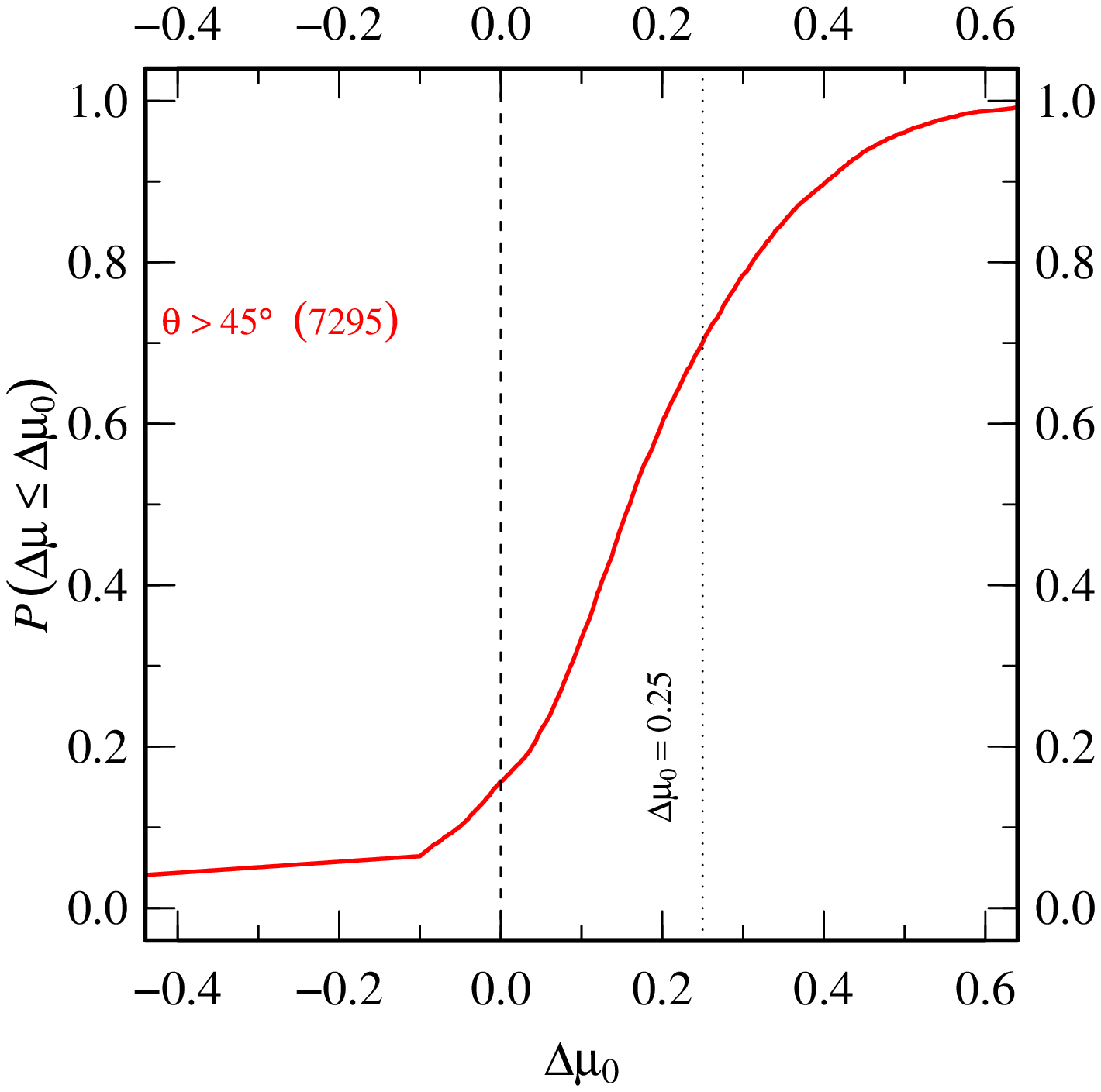}
  \caption{Left: Distribution of events in terms of fractional mass
    change and orientation change.  Middle: Cumulative fraction of
    events with spin orientation change of at least $\theta_0$, for
    major mergers ($\Delta\mu > 0.25$) and minor mergers or accretion
    ($\Delta\mu \leq 0.25$).  Right: Cumulative fraction of events
    with $\Delta\mu \leq \Delta\mu_0$, for events where the spin
    direction changes by more than $\theta_0 = 45\degr$.}
  \label{f:spinflips}
\end{figure}


\section{Spin orientation in a single object}
\begin{figure}
  \includegraphics[width=0.80\columnwidth,trim= 0 0 0 15]{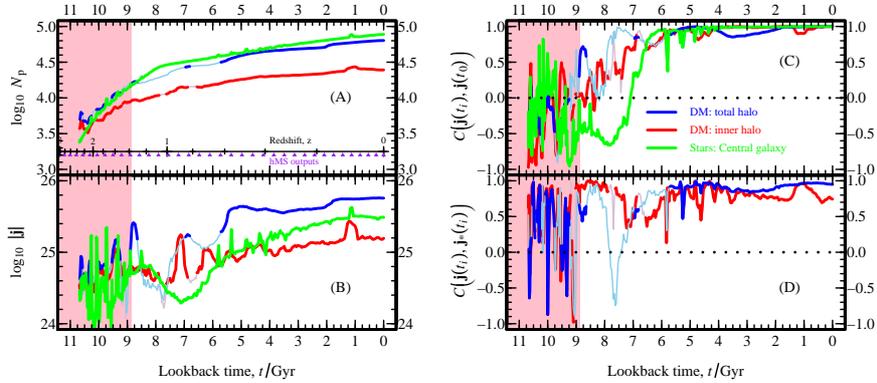}
  \caption{Evolution of a galaxy--halo system. In the pink shaded
    areas ($z\gtrsim 1.4$), the system is not yet virialised.  The
    four panels show (A) the number of particles, (B) the specific
    angular momentum magnitude ($m^2/s$), and the angular momentum orientation with
    respect to both (C) the final time and (D) the galaxy at each
    timestep, where the cosine $C(\vv{p},\vv{q}) =
    \left(\vv{p}\cdot\vv{q}\right) / \left(|\vv{p}||\vv{q}|\right)$.
    The thin lines indicate when a component fails to satisfy our
    $\jsc$ criterion, meaning its orientation is poorly defined.
  }
  \label{f:mainresults}
\end{figure}

We use a high-resolution simulation of galaxy formation in a single halo (see
\cite{Okamoto05} for details), to study the evolving relationship between halo
and galaxy spin in more detail, over many hundreds of output
times.
In Fig.~\ref{f:mainresults}, we plot the evolution of the angular
momentum magnitudes and directions of the total halo dark matter,
inner halo dark matter ($r \leq 0.25 \Rvir$), and the stellar
component of the central disc galaxy.  Without a merger occurring, the
halo flips $\sim 90\degr$ at $t\sim 8\tunit$ (although its orientation
is poorly defined at this point).  The galaxy also flips by about
$130\degr$, but over a longer timescale.  Afterwards, the galaxy and
halo are well aligned.  In this case, the disc survives, and the
``flip'' events are too slow to be seen by the analysis used in the
previous section for the hMS simulation.  In future work, we shall
relate the timescales for spin flips to the masses and other
properties of haloes and their progenitors, and characterise the
frequency of these disturbing non-merger events.

\end{document}